\documentclass[aps,prl,reprint,10pt,superscriptaddress]{revtex4-2}

\usepackage{upgreek,mathtools}
\usepackage{color}
\usepackage{graphicx}
\usepackage{nccmath}
\usepackage{bm}
\usepackage{hyperref}
\hypersetup{colorlinks,breaklinks,
            urlcolor=[rgb]{0,0,0.36},
            linkcolor=[rgb]{0,0,0.36},
            citecolor=[rgb]{0,0,0.36}}
\usepackage[mathlines]{lineno}
\usepackage{dcolumn}
\usepackage{mathtools}  
\usepackage{dsfont}
\usepackage{comment}
\usepackage{physics}
\usepackage{xcolor}
\usepackage{epsfig,tabularx,multirow,booktabs}
\usepackage{amsfonts}

\newcounter{SN}

\newcommand{\ETH}{Institute for Theoretical Physics, ETH Zurich, Wolfgang Pauli Strasse 27, 8093 Zurich, Switzerland}

\newcommand{\PKS}{Max Planck Institute for the Physics of Complex Systems, N\"{o}thnitzer Strasse 38, 01187 Dresden, Germany}

\newcommand{\Harvard}{Lyman Laboratory, Department of Physics, Harvard University, Cambridge, MA 02138, USA}

\newcommand{\Caltech}{Department of Physics and Institute for Quantum Information and Matter, California Institute of Technology, Pasadena, California 91125, USA}

\newcommand{\MIT}{Department of Physics, MIT, 77 Massachusetts Avenue, 02139 Cambridge, MA, USA}

\newcommand{\MPSD}{Max Planck Institute for the Structure and Dynamics of Matter, Luruper Chaussee 149, 22761 Hamburg, Germany}

\newcommand{\Oxford}{Clarendon Laboratory, University of Oxford, Parks Road, Oxford OX1 3PU, UK}

\begin{document}
\title{Enhanced coherence in the periodically driven two-dimensional XY model}

\author{Duilio De Santis}
\thanks{ddesantis@phys.ethz.ch}
\affiliation{\ETH}

\author{Marios H. Michael}
\affiliation{\PKS}

\author{Sambuddha Chattopadhyay}
\affiliation{\Harvard}
\affiliation{\ETH}

\author{Andrea Cavalleri}
\affiliation{\MPSD}
\affiliation{\Oxford}

\author{Gil Refael}
\affiliation{\Caltech}

\author{Patrick A. Lee}
\affiliation{\MIT}

\author{Eugene A. Demler}
\affiliation{\ETH}

\date{\today}

\begin{abstract}

Strong optical drives have been shown to induce transient superconducting-like response in materials above their equilibrium $T_c$. Many of these materials already exhibit short-range superconducting correlations in equilibrium. This motivates the question: can external driving enhance coherence in systems with superconducting correlations but no long-range order? We explore this scenario in the two-dimensional XY model with a periodically modulated stiffness using overdamped Langevin dynamics. We find that, even though the modulation leaves the average coupling unchanged, the drive can markedly increase long-range, time-averaged correlations in systems well above the equilibrium Berezinskii-Kosterlitz-Thouless temperature. The outcome depends on the ratio of the drive frequency to the intrinsic relaxation rate: faster drives primarily heat the system, suppressing correlations and conductivity. For slower drives, the optical conductivity is modified so that the real part exhibits a prolonged effective Drude scattering time, while the imaginary part has a strengthened low-frequency $1/\omega$ behavior. We map out these regimes across temperature, frequency, and amplitude, and rationalize them via simple analytics and vortex-thermalization arguments. Overall, we identify a generic nonequilibrium route to enhance coherence in XY-like systems, with potential relevance to experiments reporting light-induced superconductivity.

\end{abstract}

\maketitle

\textit{Introduction.---} The optical control of quantum materials stands at the forefront of contemporary solid-state physics, with pioneering experiments demonstrating light-induced modifications of many of the textbook phenomena in condensed matter physics: ferroelectricity, topology, and magnetism~\cite{basov_towards_2017, de_la_torre_colloquium_2021, opticalControl2, opticalControl3, opticalControl4, opticalControl5, opticalControl6, opticalControl7, opticalControl8, opticalControl9, opticalControl10, opticalControl11, floquetSolid1, floquetSolid2, floquetSolid3, floquetSolid4}. Within this program, experiments demonstrating light-induced superconducting-like response in cuprates, organic conductors, and pnictides~\cite{sc_exp1, sc_exp2, sc_exp3, sc_exp4, sc_exp5, sc_exp6, sc_exp7, sc_exp8, sc_exp9, sc_exp10, sc_exp11, sc_exp12, sc_exp13, sc_exp14,sc_exp15, sc_exp16, sc_exp17, sc_exp18, sc_exp19, sc_exp20, sc_exp21} have provided a fertile setting for broader theoretical conceptualizations of photo-driven phenomena and have galvanized intense theoretical activity~\cite{sc_theory_1, sc_theory_2, sc_theory_3, sc_theory_4, sc_theory_5, sc_theory_6, sc_theory_7, sc_theory_8, sc_theory_9, sc_theory_10, sc_theory_11, sc_theory_12, sc_theory_13, sc_theory_14}. These works have primarily focused on how strong laser driving can induce a \textit{local} superconducting order parameter above $T_c$, although other dynamical scenarios have also been proposed~\cite{sc_dynamical_scenarios1, sc_dynamical_scenarios2, sc_dynamical_scenarios3, sc_dynamical_scenarios4}.

Equilibrium measurements---for example, enhanced Nernst signals across organic and cuprate families~\cite{nernst1, nernst2, nernst3, nernst4, nernst5, nernst6}---have shown that anomalously strong superconducting fluctuations persist above $T_c$ in several materials that also exhibit light-induced superconducting-like response. These observations suggest that superconducting correlations exist locally, but the proliferation of vortices suppresses global phase coherence above $T_c$. Thus, understanding light-induced superconductivity putatively involves understanding how strongly driving a fluctuating initial state can reduce phase fluctuations and restore long-range phase coherence at temperatures far above $T_c$.

\begin{figure}
    \centering
    \includegraphics[width=\columnwidth]{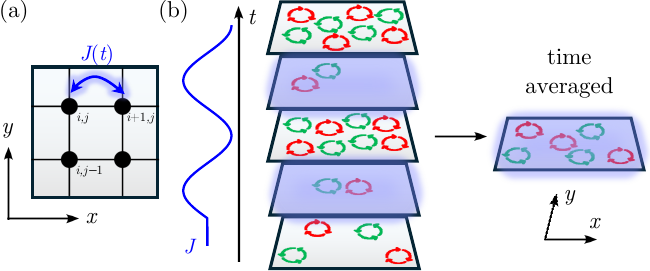}
    \caption{\textbf{Schematic of the periodically driven XY model.} (a) Square-lattice geometry of the XY model with time-modulation of the superfluid stiffness $J$ (indicated by the blue arrow). (b) Illustration of the results for for sufficiently slow driving (indicated by the blue curve). Over one drive period, when $J$ increases, long-range coherence is enhanced (shown as blue shading) and the instantaneous vortex number decreases; by contrast, when $J$ decreases, correlations are suppressed and the vortex number increases (vortices shown as green circles; antivortices as red circles). On average, the system can exhibit enhanced long-range correlations even if the time-averaged vortex density is increased compared to equilibrium.}
    \label{fig:sketch}
\end{figure}
To understand how periodic driving can restore long-range phase coherence at high temperatures, we study a driven two-dimensional (2D) XY model---paradigmatic for the Berezinskii-Kosterlitz-Thouless (BKT) scenario~\cite{Berezinskii1972DestructionLongRangeII, Kosterlitz1973OrderingMetastability, Minnhagen1987The, Jose2013FortyYearsBKT, Benfatto2024BKT}. Within overdamped noisy dynamics with a periodically modulated superfluid stiffness, see Fig.~\ref{fig:sketch}(a), we identify two regimes set by the ratio of drive frequency to intrinsic relaxation rate. When the drive is slower than relaxation, it markedly enhances long-range phase coherence: starting from a vortex-riddled state with exponentially decaying correlations, we find a strong increase of long-range time-averaged correlations, see Fig.~\ref{fig:sketch}(b). In contrast, sufficiently fast drives can heat the system and suppress correlations. We show that enhanced coherence can coincide with an increased time-averaged vortex density, highlighting the non-equilibrium character of the slow-drive regime. Computing the electromagnetic response, we find an extended effective Drude scattering time in the real part of the conductivity and an enhanced low-frequency $1/\omega$ contribution in the imaginary part, both consistent with increased coherence. We investigate these behaviors as a function of temperature, drive frequency, and amplitude. Then we provide simple analytical arguments for the observed behavior based on vortex thermalization dynamics, and finally we discuss briefly possible implications for light-induced superconducting-like responses in underdoped cuprates~\cite{Kaiser2014Optically, Hu2014OpticallyEnhanced, vonHoegen2022Amplification, Fava2024MagneticFieldExpulsion, Rosenberg2025Signatures}.

\begin{figure}
    \centering
    \includegraphics[width=\columnwidth]{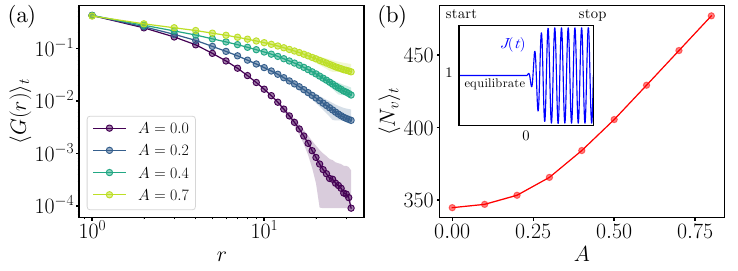}
    \caption{\textbf{Representative results versus driving amplitude.} (a) Time-averaged correlator $\langle G(r)\rangle_t$ for different driving amplitudes. Shaded areas indicate statistical errors on the average. The short-distance behavior of $\langle G(r)\rangle_t$ remains essentially unchanged from equilibrium, while long-range correlations grow markedly with $A$. (b) Time-averaged vortex-antivortex number $\langle N_v\rangle_t$ as a function of driving strength, showing a monotonic increase. Shaded areas indicate statistical errors on the average. The inset schematically shows the numerical protocol: equilibration followed by driving. The quantities in the main panels are computed only within the driving window, i.e., after equilibration at finite temperature starting from an infinite-temperature, fully random state~\cite{Supplement}. In both panels, the drive is ramped on over $n \approx 3$ periods and maintained at a steady amplitude for $\approx 10$ periods. Further parameters: $L = 64$, $T = 1.2$, $\Omega = 0.01$, $N = 10$.}
    \label{fig:driven}
\end{figure}
\textit{Model.---} We consider the 2D XY model~\cite{Berezinskii1972DestructionLongRangeII, Kosterlitz1973OrderingMetastability, Minnhagen1987The, Jose2013FortyYearsBKT, Benfatto2024BKT}
\begin{equation}
    H =  - J \sum_{\langle i,j \rangle} \cos \bigl(\theta_i - \theta_j\bigr) ,
\label{eqn:XY}
\end{equation}
describing a superconducting phase field with $\theta_i$ the phase of the complex order parameter at site $i$. Here, $J>0$ is the (in-plane) superfluid stiffness, $\langle i,j\rangle$ denotes nearest-neighbor pairs on an $L\times L$ square lattice with periodic boundary conditions, and $\theta_i\in[0,2\pi)$ is measured relative to the $x$ axis.

Our focus is on periodic modulations of the stiffness $J$, that is, we take $J \to J(t) = J_0 + f(t) A \cos(\Omega t)$, where $f(t)$ is a Gaussian envelope ensuring that the drive turns on within a few drive periods after equilibration (see below), $A$ is the drive amplitude, and $\Omega$ is the drive frequency. To investigate the resulting nonequilibrium dynamics, we equip our 2D XY framework with model-A dynamics,
\begin{equation}
    \frac{\partial \theta_i}{\partial t}
    = - \Gamma \frac{\partial H}{\partial \theta_i}
      + \eta_i(t) ,
\label{eqn:model-A}
\end{equation}
which describes overdamped relaxation, a natural choice for phase-disordered systems well above their critical temperature~\cite{Hohenberg1977Theory, Podolsky2007Nernst}. Here, $\Gamma$ is a dimensionless kinetic coefficient, and $\eta_i(t)$ is a Gaussian thermal noise satisfying $\bigl\langle \eta_i(t) \bigr\rangle = 0$ and $\bigl\langle \eta_i(t)\, \eta_j(t') \bigr\rangle = 2 \Gamma T \, \delta_{i,j} \delta(t - t')$, where $T$ is the temperature, $\delta_{i,j}$ is the Kronecker delta, and $\delta(t-t')$ is the Dirac delta. We henceforth measure time in units of $(\Gamma J_0)^{-1}$ and temperature in units of $J_0$. (To avoid notational clutter,  below we use the same symbols for the rescaled variables.) Moreover, we integrate Eqs.~\eqref{eqn:XY}--\eqref{eqn:model-A} using standard Euler and Heun finite-difference schemes, verifying the stability and consistency of our results by systematically halving the discretization steps~\cite{Ames1977Numerical, Press1992Numerical, Garcia2012Noise, Homann2024Dissipationless}. Further technical details of our treatment of the above equations, including thermalization sequences and equilibrium benchmarks, are provided in Ref.~\cite{Supplement}.
\begin{figure}
    \centering
    \includegraphics[width=0.55\columnwidth]{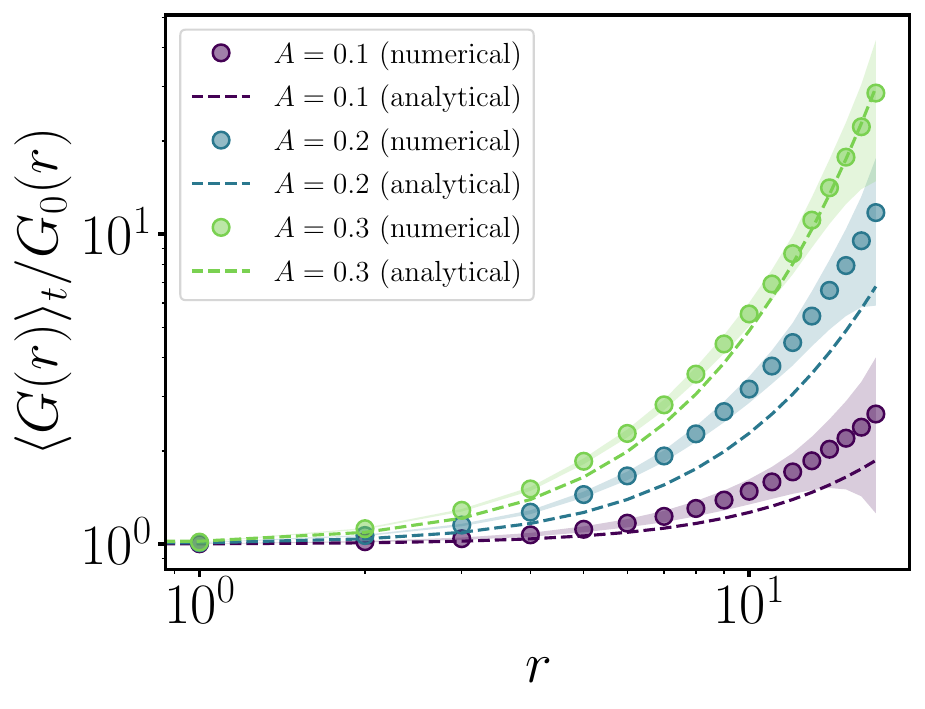}
    \caption{\textbf{Numerics versus analytics.} Ratio between the time-averaged correlator $\langle G(r)\rangle_t$ and its undriven counterpart $G_0(r)$ for different driving amplitudes. Dots indicate numerical results (shaded areas indicate statistical errors on the average); dashed lines are the analytical predictions discussed in the main text. At short distances, the ratio $\langle G(r)\rangle_t / G_0(r)$ remains close to $1$, indicating essentially unchanged correlations, while at large distances it grows markedly with $A$, reflecting enhanced correlations---a trend well captured by both numerics and analytics. Results are shown over a steady driving period. Parameters: $L = 64$, $T = 1.2$, $\Omega = 0.001$, $N = 10$.}
    \label{fig:comparison}
\end{figure}
\begin{figure*}
    \centering
    \includegraphics[width=\textwidth]{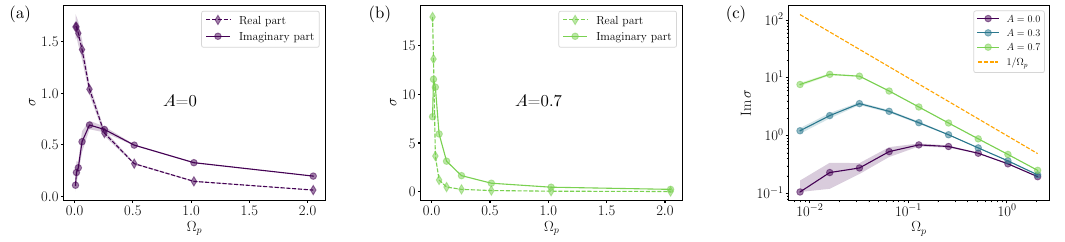}
    \caption{\textbf{Representative electromagnetic response.} (a) Real (diamonds) and imaginary (circles) parts of the conductivity versus probe frequency in the undriven case, showing the characteristic Drude-like response above $T_{\rm BKT}$. (b) Real (diamonds) and imaginary (circles) parts of the conductivity versus probe frequency under driving at $\Omega = 0.01$ and amplitude $A = 0.7$, illustrating the enhanced coherent character of the driven state. The drive is ramped on over $\approx 3$ periods and maintained at a steady amplitude for $\approx 10$ periods. (c) Log–log plot of the imaginary part of the conductivity versus probe frequency, highlighting the significantly enhanced $1/\omega$ behavior as the driving strength $A$ is increased (cf. dashed line: $1/\Omega_p$). In all panels, shaded areas indicate statistical errors on the average. Further parameters: $L = 64$, $T = 1.2$, $N = 10$, $E_{p,0} = 0.004$.}
    \label{fig:probe}
\end{figure*}
We are interested in evaluating the system's electromagnetic response and therefore include a (dimensionless) probe electric field $E_p(t)$ via a Peierls substitution with vector potential $A_p(t)$, i.e., $\cos\bigl(\theta_i - \theta_j\bigr) \to \cos\bigl(\theta_i - \theta_j - A_p\bigr)$ in Eqs.~\eqref{eqn:XY}--\eqref{eqn:model-A}, with $E_p(t) = - \partial_t A_p$~\cite{Podolsky2007Nernst}. The probe current quantifying the linear response is then given by $J_p = - \partial_{A_p} H = J(t) \sin\bigl(\theta_i - \theta_j - A_p\bigr)$. In the presence of a monochromatic probe at frequency $\Omega_p$, we Fourier transform the current to obtain $J_p(\omega)$. In the driven case, the spectrum of $J_p$ exhibits a main peak at the probe frequency $\Omega_p$ and additional satellite peaks at $\Omega_p + m\Omega$ ($m \in \mathbb{Z}$), reflecting frequency mixing between probe and drive (in contrast to the undriven equilibrium case, where no such satellites appear)~\cite{Supplement}. To quantify the linear response---the complex conductivity---we follow standard practice in ultrafast experiments and consider $\sigma = J_p(\Omega_p) / E_p(\Omega_p)$~\cite{Kaiser2014Optically, Hu2014OpticallyEnhanced, Rosenberg2025Signatures}.
\textit{Results.---} In equilibrium, we recover the standard 2D XY model, which exhibits a BKT transition from bound vortex-antivortex pairs at low temperatures to unbound vortices and antivortices at $T_{\mathrm{BKT}} \approx 0.893$~\cite{Berezinskii1972DestructionLongRangeII, Kosterlitz1973OrderingMetastability}. Accordingly, below (above) $T_{\mathrm{BKT}}$ the phase correlator $G(r = |i-j|) = \langle \cos(\theta_i - \theta_j) \rangle$, where $\langle \cdots \rangle$ denotes an equilibrium average, decays with a power law (exponentially) at sufficiently large distances. At the same time, the total number of vortices and antivortices, $N_v$, extracted from discrete circulations on the square lattice, increases sharply near the transition temperature, consistent with vortex unbinding. Details of these calculations are given in Ref.~\cite{Supplement}.

We now compute the time-averaged correlator $\langle G(r)\rangle_t$ and vortex--anti-vortex number $\langle N_v\rangle_t$ (see the inset in Fig.~\ref{fig:driven} for a sketch of the equilibration and driving protocol) under periodic modulation of the stiffness, time-averaged over several drive periods and ensemble averaged over $N$ independent realizations. As a representative case, we start from an equilibrium state with exponentially decaying correlations at $T = 1.2 > T_{\mathrm{BKT}}$, choose $\Omega = 0.01$, and scan $A$ for a system with linear size $L = 64$ and $N = 10$ runs. As the driving strength $A$ increases, the short-distance behavior of $\langle G(r) \rangle_t$ remains essentially unchanged from equilibrium, while long-range correlations grow steadily with $A$, exceeding the undriven value by orders of magnitude, see Fig.~\ref{fig:driven}(a). Thus, the drive enhances phase coherence even though the time-averaged stiffness is unchanged. We also find that this enhancement is accompanied by a monotonic increase of $\langle N_v\rangle_t$ with $A$, see Fig.~\ref{fig:driven}(b).

In the small-$\Omega$ and small-$A$ limit, these trends can be understood within a quasi-equilibrium, perturbative picture. We approximate the numerical time-averaged correlator as an average of instantaneous equilibrium correlators over a steady driving period $2\pi / \Omega$. The small driving frequency $\Omega$ ensures that thermalization is fast enough for the system to adiabatically follow the drive and develop a time-dependent correlation length $\xi(t)$. At the same time, the small-$A$ limit allows us to work with exponential correlators, as we remain close to the (above-BKT) equilibrium state. By assuming, for analytical convenience, a high-temperature form~\cite{Benfatto2024BKT} for the instantaneous correlation length, ${ \xi^{-1}(t) \approx \ln \big\lbrace 2 T / \big[ 1 + A \sin (\Omega t) \big] \big\rbrace }$, we obtain $ \langle G(r) \rangle_t \propto \frac{\Omega}{2\pi} \int_0^{2\pi / \Omega} e^{-r \, \xi^{-1}(t)} \, dt \approx G_0(r)\, I_0(r A) $, where $G_0(r)$ is the undriven correlator and $I_0$ is the modified Bessel function of the first kind~\footnote{We note that the average correlator can be evaluated using alternative expressions/inputs for $\xi^{-1}(t)$. Specifically, the analytical result is consistent with that obtained using $\xi^{-1}$ values extracted from our equilibrium numerics (not shown).}. For $r \to 0$, $I_0(r A) \sim 1 + (r A)^2/4$ yields only negligible corrections, in agreement with the numerics, whereas for large $r$ the asymptotic form $I_0(r A) \sim \exp(r A)/\sqrt{2\pi r A}$ leads to a substantial enhancement of long-range correlations, consistent with our simulations, see Fig.~\ref{fig:comparison}. A similar argument, based on the nonlinearity of the vortex–antivortex density as a function of temperature, explains the simultaneous increase in the average vortex number.

Under the driving conditions used in Fig.~\ref{fig:driven}, we now evaluate the electromagnetic probe response for different values of the driving amplitude $A$. As specified above, the conductivity $\sigma$ is defined as the ratio between the Fourier-transformed current $J_p$ at the probe frequency $\Omega_p$, induced by a weak monochromatic field $E_p(t) = E_{p,0} \sin(\Omega_p t)$ (with $E_{p,0} \ll 1$), and the probe amplitude $E_p$. In the undriven case, since our initial state lies above the BKT temperature, a Drude-like conductivity is expected, with a scattering time set by vortex collisions~\cite{Halperin1979ResistiveTransition, Ambegaokar1980Dynamics, Fisher1991Thermal}, and this behavior is indeed observed in Fig.~\ref{fig:probe}(a). Notably, as $A$ increases, Fig.~\ref{fig:probe} shows a longer Drude-like scattering time in the real part of the conductivity and an increasingly pronounced low-frequency $1/\omega$ behavior in the imaginary part, see Fig.~\ref{fig:probe}(b). The significantly enhanced $1/\omega$ behavior---corresponding to a shift of the Drude peak toward lower frequencies as the driving strength $A$ is increased---is further highlighted in Fig.~\ref{fig:probe}(c) for the imaginary part of the conductivity. These observations are consistent with enhanced coherence in the driven state. 

\begin{figure}
    \centering
    \includegraphics[width=0.65\columnwidth]{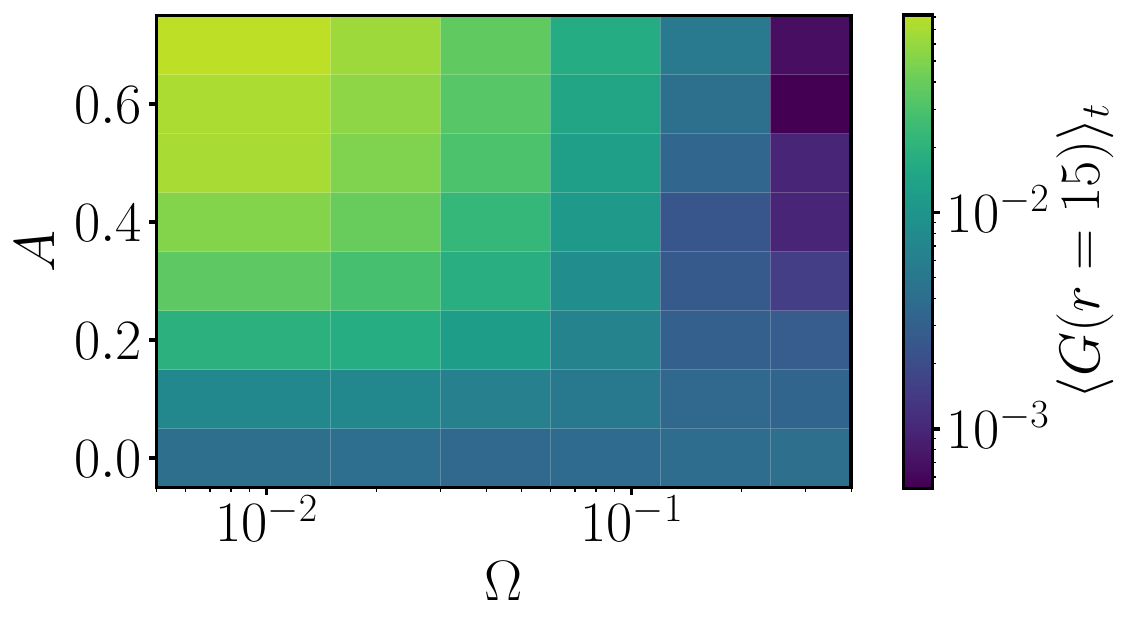}
    \caption{\textbf{Representative results versus driving frequency and amplitude.} Time-averaged correlator $\langle G(r = 15)\rangle_t$, used as a proxy for the large-distance behavior, in the $(\Omega, A)$ plane. The enhancement (light green region) is strongest in the low-$\Omega$, large-$A$ corner of parameter space, indicating that an adiabatic, strongly driven regime is most favorable for the mechanism under discussion. Here, the drive is ramped on over $n \approx 3$ periods and maintained at a steady amplitude for $\approx 10$ periods. Further parameters: $L = 64$, $T = 1.2$, $N = 10$.}
    \label{fig:freq-scan}
\end{figure}
So far, we have considered the limit of small driving frequencies. In the large-$\Omega$ limit, by contrast, the system becomes sensitive only to the time average of the drive, which is zero in our case, so the enhancement is expected to eventually vanish. To investigate this, we perform a sweep in the $(\Omega, A)$ plane and use $\langle G(r = 15)\rangle_t$ as a proxy for the large-distance correlation behavior. As shown in Fig.~\ref{fig:freq-scan}, the correlation enhancement monotonically decreases with increasing $\Omega$, with signatures of heating and correlation suppression at the highest driving amplitudes (see the purple region). Overall, the enhancement (see the light green region) is strongest in the low-$\Omega$, large-$A$ corner of parameter space, indicating that an adiabatic, strongly driven regime is most favorable for the mechanism under discussion. It is worth noting, however, that conductivities with a significantly enhanced coherent character---though reduced relative to the adiabatic limit---are still observed at higher driving frequencies, see the Supplementary Material~\cite{Supplement}.

The above behavior is natural in our overdamped dynamics, where the bottleneck for developing power-law-like long-range correlations is the thermalization of vortex excitations during the nonequilibrium sequence. In the Supplementary Material, see Ref.~\cite{Supplement}, we analyze this in more detail using quench protocols that mimic segments of the sinusoidal drive. There we discuss that the characteristic growing length (roughly, the intervortex distance) scales as $\sqrt{t / \ln t}$ following a quench to a larger value of $J$~\cite{Yurke1993CoarseningXY, Bray1994GrowthLaws, Rutenberg1995EnergyScaling, Jelić2011Quench}. As $\Omega$ increases, the time spent in the favorable high-stiffness part of each cycle becomes too short for this length scale to grow sufficiently. Short-distance correlations can still adjust quickly, but long-range correlations (e.g., at $r = 15$) remain underdeveloped in time and, consequently, at the time-averaged level, in agreement with our observations.

As a further remark, we note that the strength of these effects is also temperature dependent. Further above $T_{\mathrm{BKT}}$, while keeping the representative driving parameters used before, thermalization within each drive cycle becomes faster. This can, in turn, favor an increase of the average correlation length at relatively higher drive frequencies compared to the near-critical regime, see the Supplementary Material~\cite{Supplement}. In this parameter range, enhanced correlations may even arise without a corresponding increase in the vortex density relative to equilibrium.
\textit{Discussion and outlook.---} We have shown that periodic modulation of the phase stiffness in overdamped XY models generically promotes long-range phase coherence above the critical temperature. In a slow-drive regime, where the system can follow the modulation within each cycle, the drive amplifies phase coherence. A configuration that is initially vortex-riddled and has only short-range correlations develops enhanced time-averaged correlations that extend over substantially longer distances. An interesting outcome is that stronger coherence does not require vortex depletion: in fact, the slow-drive regime can exhibit an increased average density of vortices, underscoring its nonequilibrium character. The analyzed electromagnetic response reflects the induced coherence through a lengthened effective Drude scattering time in $\mathrm{Re}\,\sigma(\omega)$ and a stronger $1/\omega$-like low-frequency component in $\mathrm{Im}\,\sigma(\omega)$. Altogether, our results establish a generic nonequilibrium route to enhancing the long-range coherence of a strongly phase-fluctuating state via external periodic driving, broadly applicable to systems in the XY universality class.

In particular, our ideas may have implications for the theoretical understanding of recent reports of light-induced superconducting-like behavior well above the equilibrium transition temperature, observed in pump--probe experiments that selectively drive phonons in underdoped cuprates~\cite{Kaiser2014Optically, Hu2014OpticallyEnhanced, vonHoegen2022Amplification, Fava2024MagneticFieldExpulsion, Rosenberg2025Signatures}. Much of the earlier data involve out-of-plane response due to a drive field polarized perpendicular to the plane, and theoretical discussions have been provided for this setting~\cite{Michael2025Parametrically}. However, recently an interesting in-plane response has been reported~\cite{Rosenberg2025Signatures}. An accurate description of this experiment requires discussing the microscopic mechanisms by which electromagnetic pulses polarized perpendicular to the planes generate stiffness modulation, the role of YBCO’s bilayer structure, and the relative timescales of the pump pulse and phase-relaxation dynamics. These questions will be addressed in a separate publication~\cite{DeSantis2025}.
%

%
We thank S.A. Kivelson, A. Gómez Salvador, J. Curtis, R. Andrei, O. Malyshev, M. Buzzi, D. Nicoletti, M. Rosenberg, P.E. Dolgirev, and S. Sachdev for insightful discussions. D.D.S. acknowledges support from the “Angelo Della Riccia” Foundation and ETH Zurich. P.A.L. acknowledges support by DOE (USA) office of Basic Sciences Grant No. DE-FG02-03ER46076. E.D. acknowledges the Swiss National Science Foundation (Grant Number 200021\textunderscore212899) and ETH-C-06 21-2 equilibrium Grant with project number 1-008831-001 for funding.
%


\bibliographystyle{apsrev4-2}
\bibliography{refs}


\appendix
\clearpage
\newpage
\setcounter{figure}{0}
\setcounter{equation}{0}
\setcounter{page}{1}

\renewcommand{\thepage}{S\arabic{page}} 
\renewcommand{\thesection}{S\arabic{section}} 
\renewcommand{\thetable}{S\arabic{table}}  
\renewcommand{\thefigure}{S\arabic{figure}} 
\renewcommand{\theequation}{S\arabic{equation}} 

\onecolumngrid

\begin{center}
\textbf{\large{Supplemental Material \\ Enhanced coherence in the periodically driven two-dimensional XY model}}
\end{center}

\section{Numerical details}
\label{sec:numerics}

We consider the classical XY model on a square lattice with spacing $a$ and phases $\{\theta_i\}$, $\theta_i \in [0,2\pi)$, with Hamiltonian
\begin{equation}
    H[\{\theta\}]
    = -J \sum_{\langle ij \rangle} \cos(\theta_i - \theta_j),
\end{equation}
where $J$ is the nearest-neighbor coupling (taken time independent here, $J \equiv J_0$, for simplicity) and $\langle ij \rangle$ denotes nearest-neighbor pairs. We impose periodic boundary conditions in both spatial directions.

The system evolves under purely relaxational (model-A) Langevin dynamics for a non-conserved phase field:
\begin{equation}
    \frac{d \theta_i}{dt}
    = -\Gamma \frac{\partial H}{\partial \theta_i} + \eta_i(t),
\end{equation}
with kinetic coefficient $\Gamma$ and Gaussian white noise $\eta_i(t)$ satisfying
\begin{align}
    \langle \eta_i(t) \rangle &= 0, \\
    \langle \eta_i(t)\,\eta_j(t') \rangle &= 2 \Gamma T \,\delta_{ij}\,\delta(t-t'),
\end{align}
where $T$ is the temperature~\cite{Hohenberg1977Theory, Podolsky2007Nernst}.

Using
\begin{equation}
    \frac{\partial H}{\partial \theta_i}
    = J \sum_{j \in nn(i)} \sin(\theta_i - \theta_j),
\end{equation}
the Langevin equation becomes
\begin{equation}
    \frac{d \theta_i}{dt}
    = -\Gamma J \sum_{j \in nn(i)} \sin(\theta_i - \theta_j)
      + \eta_i(t).
    \label{eq:langevin-xy}
\end{equation}

\textit{Euler discretization.---} Let the discrete time be $t_n = n \,\delta t$ with time step $\delta t$. We define the deterministic drift
\begin{equation}
    f_i(\{\theta\})
    = -\Gamma J \sum_{j \in nn(i)} \sin(\theta_i - \theta_j).
\end{equation}
The Euler update for Eq.~\eqref{eq:langevin-xy} is
\begin{equation}
    \theta_i^{n+1}
    = \theta_i^n
      + f_i(\{\theta^n\})\,\delta t
      + \sqrt{2 \Gamma T \,\delta t}\;\xi_i^n,
\end{equation}
where $\xi_i^n$ are independent standard normal variables~\cite{Garcia2012Noise},
\begin{equation}
    \langle \xi_i^n \rangle = 0, \qquad
    \langle \xi_i^n \xi_j^m \rangle = \delta_{ij}\,\delta_{nm}.
\end{equation}
After each step one may take $\theta_i^{n+1} \mapsto \theta_i^{n+1} \bmod 2\pi$.

\textit{Heun (predictor-corrector) discretization.---} For additive noise, the Heun scheme improves the deterministic part
while using the same noise realisation in predictor and corrector.

\begin{enumerate}
    \item Predictor step:
    \begin{equation}
        \tilde{\theta}_i
        = \theta_i^n
          + f_i(\{\theta^n\})\,\delta t
          + \sqrt{2 \Gamma T \,\delta t}\;\xi_i^n.
    \end{equation}
    \item Corrector step:
    \begin{equation}
        \theta_i^{n+1}
        = \theta_i^n
          + \frac{\delta t}{2}
            \left[ f_i(\{\theta^n\})
                 + f_i(\{\tilde{\theta}\}) \right]
          + \sqrt{2 \Gamma T \,\delta t}\;\xi_i^n.
    \end{equation}
\end{enumerate}

Again, angles may be taken modulo $2\pi$ after each update.

We show numerical runs for the Euler scheme, with typical $\delta t \cdot (\Gamma J) = 0.05 - 0.01$. As in the main text, we henceforth measure time in units of $(\Gamma J)^{-1}$ and temperature in units of $J$, where $J$ is understood as the equilibrium value. To avoid notational clutter, we use the same symbols for the dimensionless variables.

\section{Equilibrium benchmark}
\label{sec:benchmark}

In the absence of driving, our model reduces to the conventional 2D XY system on a square lattice, characterized by a BKT transition at $T_{\mathrm{BKT}} \approx 0.893$~\cite{Berezinskii1972DestructionLongRangeII, Kosterlitz1973OrderingMetastability}. Below this temperature, vortices and antivortices predominantly form bound pairs and the phase correlator $G(r) = \langle \cos(\theta_i - \theta_j) \rangle$ exhibits algebraic decay, whereas above $T_{\mathrm{BKT}}$ free vortices proliferate and $G(r)$ instead decays exponentially at large separations.

\begin{figure}
    \centering
    \includegraphics[width=0.9\textwidth]{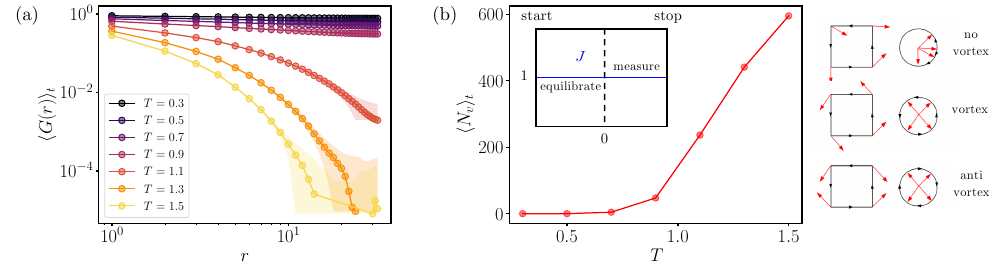}
    \caption{\textbf{Equilibrium behavior versus temperature.} (a) $G(r)$ correlator for different temperatures. A crossover from power-law to exponential behavior is observed around $T \approx 0.9$. (b) $N_v$, the total vortex–antivortex number as a function of temperature. This quantity increases sharply near the transition temperature, consistent with vortex unbinding. The inset schematically shows the equilibration protocol; the quantities in the main panels are computed only within the measurement window of the simulations. The sketches on the right show the discrete circulations on the square lattice used to compute the vorticity (red arrows represent phase variables at lattice sites). In all panels, shaded areas indicate statistical errors on the average. Parameters: $L = 64$, $N = 10$, $\Delta t_b \approx \Delta t_m \approx 10^4$.}
    \label{fig:EQ}
\end{figure}

Here we describe how we prepare the initial equilibrium state. We start our simulations from a fully random, infinite-temperature configuration, then equilibrate at a chosen finite temperature $T$ over a burn-in interval $\Delta t_{b}$. The required $\Delta t_{b}$ depends on temperature and system size, and is determined by monitoring the relaxation of observables such as the energy per site and the magnetization. After equilibration, we measure the quantities of interest over a time window $\Delta t_{m}$, so that the total simulated time is $\Delta t_{s} = \Delta t_{b} + \Delta t_{m}$, see the inset in Fig.~\ref{fig:EQ}(b) for a sketch. For representative parameter sets and averaging over an ensemble of $N$ independent realizations, our relaxational dynamics reproduces the expected BKT phenomenology: a crossover of $G(r)$ from power-law to exponential behavior around $T \approx 0.9$, see Fig.~\ref{fig:EQ}(a). We also compute the vorticity field, defined via discrete circulations of the $\theta$ field on the square lattice, extract the total number of vortices and antivortices $N_v$, and find that it increases sharply near the transition temperature, consistent with vortex unbinding; see Fig.~\ref{fig:EQ}(b).

\section{Probing protocol}
\label{sec:probe}

To probe the system's electromagnetic response, we apply a weak monochromatic electric field $E_p(t) = E_{p,0} \sin(\Omega_p t)$, switched on only after equilibration (see previous section). The coupling to the field is implemented via a Peierls substitution with vector potential $A_p(t)$, so that $ \cos(\theta_i - \theta_j) \;\to\; \cos\bigl(\theta_i - \theta_j - A_p\bigr) $, with $E_p(t) = -\partial_t A_p(t)$. We choose $E_{p,0} \ll 1$, small enough to ensure that the probe perturbs the system much more weakly than the typical drive, yet large enough to yield a clear signal; its value is systematically doubled to verify linearity of the response. Furthermore, we measure over more than 10 full probe periods, and we emphasize that linear response data at different probe frequencies $\Omega_p$ are obtained from independent simulations, each performed with a single monochromatic probe.

\begin{figure}
    \centering
    \includegraphics[width=0.75\textwidth]{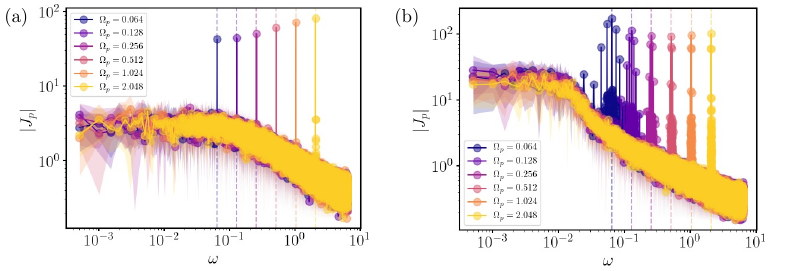}
    \caption{\textbf{Probe current spectra: equilibrium versus driven.} (a) Equilibrium ($T > T_{\rm BKT}$) $|J_p(\omega)|$ spectrum versus frequency for different probe frequencies $\Omega_p$ (indicated by dashed vertical lines). In all cases, the response is sharply peaked at $\omega = \Omega_p$. (b) Driven $|J_p(\omega)|$ spectrum versus frequency for different probe frequencies $\Omega_p$ (dashed vertical lines). The drive, with frequency $\Omega = 0.01$ and amplitude $A = 0.7$, is ramped on over $\approx 3$ periods and maintained at a steady amplitude for $\approx 10$ periods. In addition to the main peak at $\omega = \Omega_p$, the spectrum now exhibits sidebands at $\Omega_p + m\Omega$, reflecting frequency mixing between the probe and the drive. Parameters: $L = 64$, $T = 1.2$ $N = 10$, $E_{p,0} = 0.004$.}
    \label{fig:probe_1col}
\end{figure}

The corresponding probe current is $J_p = -\partial_{A_p} H = J(t)\,\sin\bigl(\theta_i - \theta_j - A_p\bigr)$, from which we obtain $J_p(\omega)$ by Fourier transformation. In equilibrium (without driving), $|J_p(\omega)|$ displays a single peak at the probe frequency $\Omega_p$, as illustrated in Fig.~\ref{fig:probe_1col}(a) for a representative probe-frequency scan above $T_{\rm BKT}$. Once the periodic drive is present, the spectrum develops additional sidebands at $\Omega_p + m\Omega$ ($m \in \mathbb{Z}$), reflecting frequency mixing between the probe and the drive, see Fig.~\ref{fig:probe_1col}(b). Following standard practice in pump--probe studies of driven cuprates~\cite{Kaiser2014Optically, Hu2014OpticallyEnhanced, Rosenberg2025Signatures}, we characterize the linear response via the complex conductivity $\sigma(\Omega_p) = J_p(\Omega_p) / E_{p,0}$. The data in Fig.~\ref{fig:probe_1col} are therefore used to construct the results shown in Fig.~\ref{fig:probe} in the main text.

\section{Electromagnetic response vs. driving frequency $\Omega$}
\label{sec:more_probe}

\begin{figure*}
    \centering
    \includegraphics[width=\textwidth]{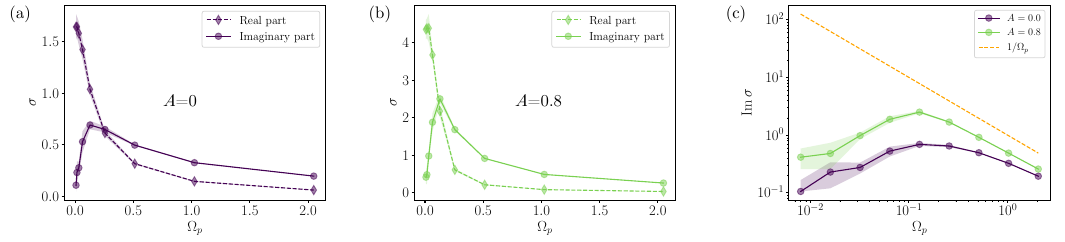}
    \caption{\textbf{Equilibrium versus driven conductivities.} (a) Real (diamonds) and imaginary (circles) parts of the conductivity versus probe frequency in the undriven case, exhibiting the characteristic Drude-like response above $T_{\rm BKT}$. (b) Real (diamonds) and imaginary (circles) parts of the conductivity versus probe frequency under periodic driving at $\Omega = 0.1$ and amplitude $A = 0.8$, showing a more coherent response: a longer effective Drude time in $\mathrm{Re}\,\sigma$ and an enhanced low-frequency tail in $\mathrm{Im}\,\sigma$. The drive is ramped on over $\approx 3$ periods and then held at a steady amplitude for $\approx 50$ periods. (c) Log--log plot of $\mathrm{Im}\,\sigma$ versus probe frequency, emphasizing the strengthened $1/\omega$ behavior with increasing $A$ (cf. dashed line: $1/\Omega_p$). Further parameters: $L = 64$, $T = 1.2$, $N = 10$, $E_{p,0} = 0.004$.}
    \label{fig:probe_supp}
\end{figure*}

Here we analyze conductivity results for a drive frequency $\Omega = 0.1$, i.e., an order of magnitude higher than in Fig.~\ref{fig:probe} of the main text. In the undriven case, because the initial state lies above the BKT temperature, a Drude-like response is expected with a scattering time set by vortex collisions; see Fig.~\ref{fig:probe_supp}(a). Under driving, increasing $A$ yields a longer effective Drude scattering time in $\mathrm{Re}\,\sigma(\omega)$ and a more pronounced low-frequency $1/\omega$ contribution in $\mathrm{Im}\,\sigma(\omega)$; see Fig.~\ref{fig:probe_supp}(b). The strengthened $1/\omega$ behavior is further highlighted in Fig.~\ref{fig:probe_supp}(c). Although these enhancements are smaller than those at lower drive frequency [cf. Fig.~\ref{fig:probe} in the main text], they remain indicative of increased coherence in the driven state, are consistent with recent in-plane measurements on driven underdoped cuprates~\cite{Rosenberg2025Signatures}, and underscore the robustness of the mechanism.

\section{Quench dynamics}
\label{sec:quench}

\begin{figure*}
    \centering
    \includegraphics[width=\textwidth]{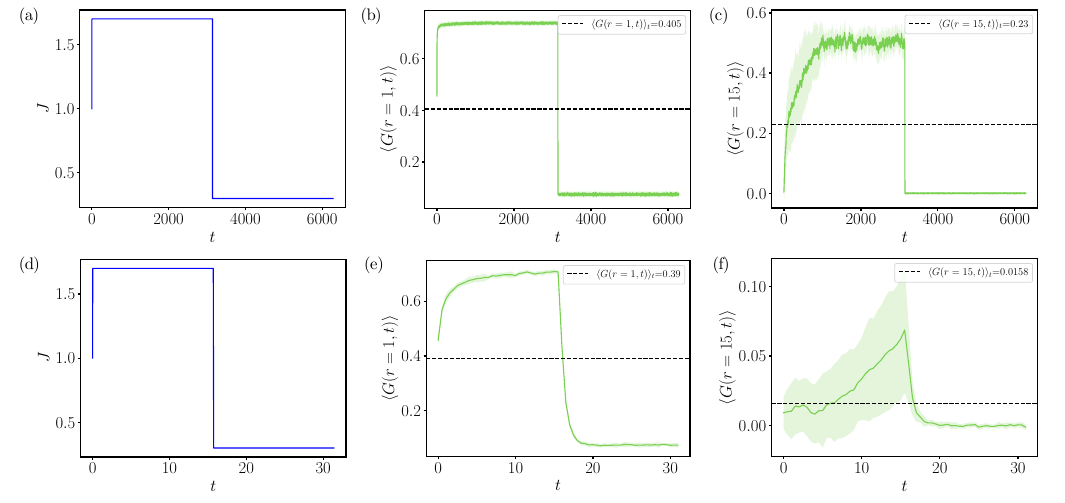}
    \caption{\textbf{Time-resolved correlators for quenches over different timescales.} (a) [(d)]  Quench profile over a timescale of $2\pi/\Omega$, with $\Omega = 0.001$ [$\Omega = 0.2$]. (b) [(e)] Time-resolved correlators at $r = 1$. (c) [(f)] Time-resolved correlators at $r = 15$. Shaded areas indicate statistical errors on the ensemble average, and the dashed horizontal lines mark the corresponding time averages. While short-distance correlations at $r = 1$ [panels (b) and (e)] are comparable for the two quenches, at larger distances $r = 15$ [panels (c) and (f)] the faster quench leads to markedly weaker correlations. Further parameters: $L = 64$, $T = 1.2$, $A = 0.7$, $N = 10$.}
    \label{fig:quench}
\end{figure*}

\begin{figure*}
    \centering
    \includegraphics[width=0.65\textwidth]{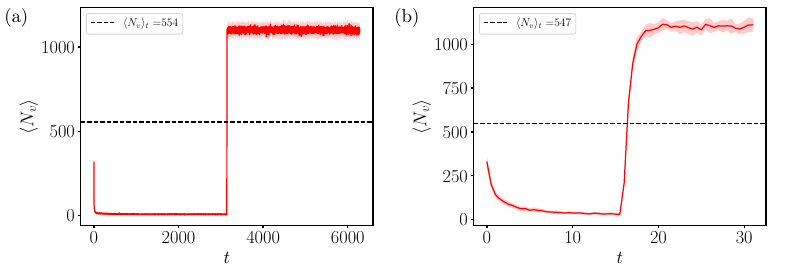}
    \caption{\textbf{Instantaneous vortex number for quenches on different timescales.} (a) [(b)] Time trace of the vortex–antivortex count under a quench of duration $2\pi/\Omega$ with $\Omega=0.001$ [$\Omega=0.2$]. Shaded areas indicate ensemble-averaged statistical uncertainties; dashed horizontal lines denote the corresponding time averages. In both cases, the nonlinear dependence of vortex dynamics on the coupling permits the time-averaged vortex number to exceed its equilibrium value. Further parameters: $L = 64$, $T = 1.2$, $A = 0.7$, $N = 10$.}
    \label{fig:quench_vortex}
\end{figure*}

Here we discuss the thermalization of vortex excitations during the nonequilibrium sequence. To gain insight into this process, we consider a quench protocol that mimics one period of the sinusoidal drive at frequency $\Omega$ discussed in the main text: a quench to a high-effective-stiffness state held for some time, followed by a quench to a low-effective-stiffness state for the remainder of the cycle. Long-range correlations in $\langle G(r)\rangle_t$ predominantly build up during the high-stiffness interval, but their growth relies on slow vortex collisions. It is known that, after an infinite-temperature quench to a temperature below $T_{\mathrm{BKT}}$, the so-called growing length (roughly, the intervortex distance) behaves as $\sqrt{t / \ln t}$~\cite{Yurke1993CoarseningXY, Bray1994GrowthLaws, Rutenberg1995EnergyScaling, Jelić2011Quench}. As $\Omega$ increases, the time spent in the favorable high-stiffness segment of each cycle becomes insufficient for correlations at large $r$ to fully develop, and the enhancement over equilibrium is progressively lost, as anticipated from this nearly diffusive scaling law. We illustrate this in Fig.~\ref{fig:quench}, where panels (a)–(c) [(d)–(f)] correspond to a quench over a timescale of $2\pi/\Omega$, with $\Omega = 0.001$ [$\Omega = 0.2$]. The driving amplitude is fixed at $A = 0.7$. While short-distance correlations at $r = 1$ [panels (b) and (e)] evolve similarly in time for both quenches, the behavior at larger distances, $r = 15$ [panels (c) and (f)], differs dramatically: the faster quench exhibits significantly underdeveloped correlations in time and, consequently, at the time-averaged level, confirming our earlier observations. In both the sinusoidal-drive and quench scenarios, the nonlinearity of the vortex dynamics with respect to the coupling allows the time-averaged vortex number to exceed its equilibrium value, see Fig.~\ref{fig:quench_vortex}, as noted in the main text.

\section{Time-averaged correlations vs. temperature $T$}
\label{sec:highT}

\begin{figure}
    \centering
    \includegraphics[width=0.35\textwidth]{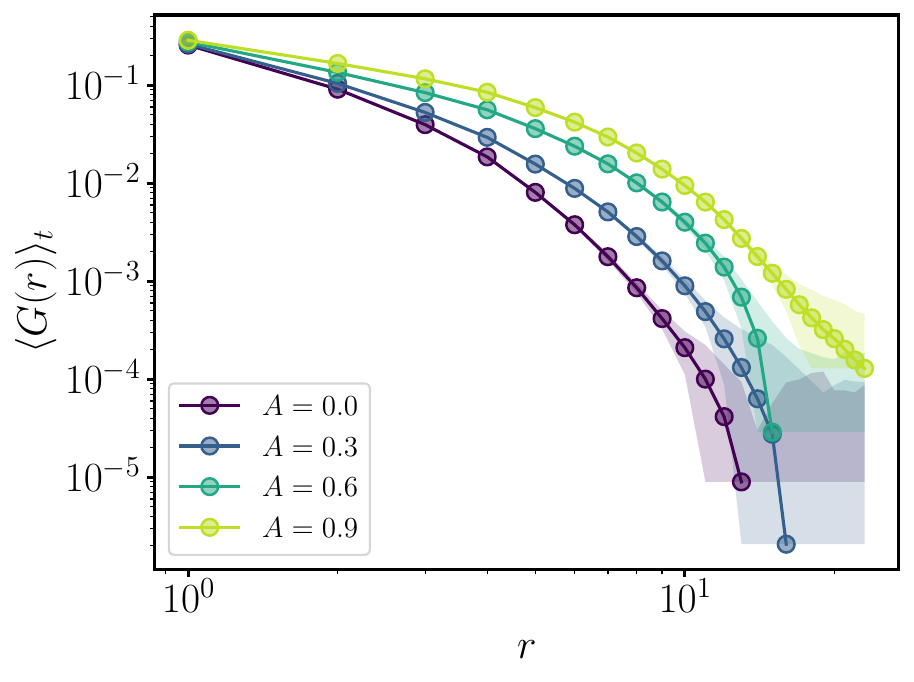}
    \caption{\textbf{Representative results at higher initial temperature.} (a) Time-averaged correlator $\langle G(r)\rangle_t$ for different driving amplitudes. Shaded areas indicate statistical errors on the average. We observe an increase of the average correlation length at relatively higher drive frequencies compared to that in the near-critical regime, see Fig.~\ref{fig:driven} in the main text. The drive is ramped on over $n \approx 3$ periods and maintained at a steady amplitude for $\approx 10$ periods. Further parameters: $L = 64$, $T = 1.6$, $\Omega = 0.2$, $N = 10$.}
    \label{fig:highT}
\end{figure}

The magnitude of the drive-induced effects also depends on the temperature $T$. Further above $T_{\mathrm{BKT}}$ than in the main text, while keeping representative driving parameters, intra-cycle thermalization becomes faster. This, in turn, can favor growth of the average correlation length (see Fig.~\ref{fig:highT}) at drive frequencies that are comparatively higher than in the near-critical regime. In this regime, we may observe enhanced correlations without a corresponding increase in the vortex density relative to equilibrium (not shown).

\end{document}